\documentclass[reprint, superscriptaddress, amsmath, amssymb, aps, pra]{revtex4-2}

\usepackage{graphicx}
\usepackage{dcolumn}
\usepackage{bm}
\usepackage{algorithm2e}
\usepackage{alltt}
\usepackage{braket}
\usepackage[colorlinks=true,linkcolor=blue,urlcolor=blue,citecolor=blue,pdfusetitle]{hyperref}

\usepackage{tikz}
\usepackage{lipsum}

\RestyleAlgo{ruled}
\SetKwInput{Kw}{input}
\SetKwData{Precompute}{Precompute}
\SetProgSty{} 

\newcommand{\tr}{\mathrm{tr}}
\newcommand{\ketbra}[2]{\left| #1 \vphantom{#2} \right>\!\! \left< #2 \vphantom{#1} \right|}
\DeclareMathOperator{\vectorize}{Vec}
\DeclareMathOperator{\unvectorize}{Unvec}

\begin{document}

\title{Gillespie algorithm for quantum jump trajectories}

\author{Marco Radaelli}
\email{radaellm@tcd.ie}
\affiliation{School of Physics, Trinity College Dublin, Dublin 2, Ireland}
\affiliation{Trinity Quantum Alliance, Unit 16, Trinity Technology and Enterprise Centre, Pearse Street, Dublin 2, D02 YN67, Ireland}

\author{Gabriel T. Landi}
\email{gabriel.landi@rochester.edu}
\affiliation{Department of Physics and Astronomy, University of Rochester, Rochester, New York 14627, USA}

\author{Felix C. Binder}
\email{quantum@felix-binder.net}
\affiliation{School of Physics, Trinity College Dublin, Dublin 2, Ireland}
\email{quantum@felix-binder.net}
\affiliation{Trinity Quantum Alliance, Unit 16, Trinity Technology and Enterprise Centre, Pearse Street, Dublin 2, D02 YN67, Ireland}

\begin{abstract}
The jump unravelling of a quantum master equation decomposes the dynamics of an open quantum system into abrupt jumps, interspersed by periods of coherent dynamics when no jumps occur. Such open quantum systems are ubiquitous in quantum optics and mesoscopic physics, hence the need for efficient techniques for their stochastic simulation.
Numerical simulation techniques fall into two main categories. The first splits the evolution into small timesteps and determines stochastically for each step if a jump occurs or not. The second, known as Monte Carlo Wavefunction simulation, is based on the reduction of the norm of an initially pure state in the conditional no-jump evolution. It exploits the fact that the purity of the state is preserved by the finest unraveling of the master equation.
In this work, we present an alternative method for the simulation of the quantum jump unraveling, inspired by the classical Gillespie algorithm. The method is particularly well suited for situations in which a large number of trajectories is required for relatively small systems. It allows for non-purity-preserving dynamics, such as the ones generated by partial monitoring and channel merging. 
We describe the algorithm in detail and discuss relevant limiting cases. To illustrate it, we include four example applications of increasing physical complexity and discuss the performance of the algorithm across regimes of interest for open quantum systems simulation. 
Publicly available implementations of our code are provided in Julia and Mathematica. 
\end{abstract}

\maketitle

\section{Introduction}

Quantum master equations have become an absolutely essential methodology for most areas of quantum physics. They are used to describe experiments in a wide variety of platforms, from quantum optics to mesoscopic electronics. 
The quintessential Gorini–Kossakowski–Sudarshan–Lindblad (GKSL) master equation has the form~\cite{Lindblad1976,Gorini1976, breuer2007,schaller2014}
\begin{equation}\label{M}
    \frac{d\rho}{dt} = \mathcal{L}\rho = -i [H,\rho] + \sum_{k=1}^r \mathcal{D}[L_k] \rho,
\end{equation}
where $\mathcal{L}$ is the Liouvillian, 
$\rho$ the state of the system, 
$H$ the Hamiltonian, and $\mathcal{D}[L]\bullet = L \bullet L^\dagger - \frac{1}{2} \{L^\dagger L, \bullet\}$ a Lindblad dissipator.

While the master equation describes the ensemble-averaged evolution of the system's density matrix $\rho(t)$, one can also \emph{unravel} it in terms of specific quantum trajectories~\cite{Diosi1985, Javanainen1986, Diosi_1988, Plenio1998, Wiseman_2009,Carmichael2000}. 
The quantum jump unravelling (QJU), in particular, separates the dynamics into a stochastic process consisting of abrupt jumps occurring at random times, with a unitary no-jump evolution in between. 
The jumps are associated with the terms~$L_k\rho L_k^\dagger$, each representing a possible ``jump channel''. 
The QJU method has been extensively employed in various contexts, for many decades.
Its main motivation lies in the fact that, in many experimentally relevant systems, the quantum jumps are directly associated with clicks in specific detectors, as is the case, for instance, for photon-detection in optical systems. Here, the stochastic dynamics can be used to study the emission spectrum~\cite{Marte_1993}, or to obtain the full counting statistics of photo-detection~\cite{Kewming2022}.
A beautiful recent example is the experiment reported in Ref.~\cite{Fink_2018}, which used photo-detection stochastic trajectories to build up the statistics necessary to demonstrate a driven-dissipative phase transition. 
Similarly, in mesoscopic physics, the QJU can be used to model coherent electron tunneling from quantum dots to metallic leads~\cite{Goan_2001_1,schaller2014,Landi2022}. 
In this case, the  jump channels correspond to the injection/extraction of an electron onto/from the quantum dot.
Finally, the QJU can also be used as a method to simulate the solution $e^{\mathcal{L}t} \rho(0)$ of Eq.~\eqref{M}.

Such a stochastic simulation is formally equivalent to what is often referred to as \emph{Monte Carlo Wavefunction} (MCW) simulation~\cite{Cohen_1986, Dalibard_1992, Molmer_1996, Dum_1992}. It exploits the reduction of the norm of the initial (pure) state under conditional no-jump dynamics as a means to effectively sample the waiting time distribution between jumps. The power of MCW simulation lies in the fact that it allows one to work with pure states which are numerically much easier to handle than mixed states in systems with high-dimensional Hilbert spaces due to correspondingly lower memory requirements~\cite{gardiner2004}. In particular, the finest unraveling of a master equation preserves purity; if the initial state is mixed, it is sufficient to consider its decomposition as a convex mixture of pure states and to evolve each component separately. Due to the linearity of quantum evolution, every property of the mixed state can be obtained as the same convex combination of the properties of each pure state value.

There are, however, situations where working with pure states is not an option. An important case are systems subject to partial monitoring, i.e., when only a subset of the jump channels is monitored, as well as cases where emissions in some of the channels cannot be distinguished one from each other (we will refer to this as \textit{channel merging} in the following). In these scenarios, purity is, in fact, not preserved along the evolution, and one can choose between two main methods.

Here, one could use the full-monitoring master equation to generate trajectories via the MCW method. Such trajectories could then be post-processed to obtain partial monitoring or channel merging. However, in order to faithfully sample the reduced dynamics, one would need a number of fully monitored trajectories that scales exponentially with time (for partial monitoring) or number of jumps (for channel merging).

Alternatively, we can directly evolve the density matrix of the system with the partial monitoring GKSL equation. While the evolution of the entire density matrix is more computationally expensive than the evolution of the state vector alone, this is compensated by the reduction in the number of required trajectories. A common method to evolve the density matrix is outlined in Sect.~\ref{sec:QJU}, and is based on the random selection, at each timestep, of whether an emission should or should not occur in that timestep. Given its stepwise nature, we refer to this method as step-by-step QJU (SS-QJU) in the following. 

The main shortcoming of the SS-QJU method is that it may require very small time steps for the integration of the stochastic master equation to converge. 
This happens for the following reason. 
In quantum coherent systems, the dynamics are always a mixture of the unitary dynamics, which cause no jumps, and dissipative terms responsible for jumps.
The difficulty is when the unitary dynamics occur at a much shorter timescale, forcing one to use very small timesteps, even though the jumps occur on a much longer timescale. 

A similar problem is also found in classical master equations evolving under different timescales. 
In that field, the Gillespie algorithm~\cite{Gillespie_1976, Gillespie_1977} stands out as an extremely efficient alternative method for simulating the behaviour of such dissipative dynamics. 
The algorithm is based on the Waiting-Time Distribution (WTD) $W(t)$, which describes the time it takes until the next jump occurs~\cite{brandes_2008,Landi2023}. 
Instead of integrating an equation over small time steps, the Gillespie algorithm samples a random time $T$ from $W(t)$. This represents the time for the next jump will occur. 
Next, it  randomly selects one of the possible jump channels. 
In this fashion, the system is directly propagated forward in time, to the configuration after the jump. 
In classical master equations $W(t)$ is always exponentially distributed, making it very easy to sample the random times $T$. 
In addition, classical master equations belong to the class of \emph{renewal processes}: after the jump the state of the system is completely reset, losing any possible memory of previous configurations. As a consequence, $W(t)$ only depends on the time elapsed since the most recent jump of the system, and not on the system's previous history. 

In quantum systems, these two features are generally absent: the unitary part of the dynamics is responsible for WTDs which are not exponentially distributed; and jumps do not fully reset the state of the system in general. 
Nonetheless, as we show in this paper, a quantum version of the Gillespie algorithm can be derived and implemented, and indeed results in an efficient simulation method.
In particular, the quantum Gillespie algorithm we implement has a moderate computational overhead in terms of the quantities that must be pre-computed (at the cost of memory requirement). Once this is done, the simulation of the quantum trajectories becomes comparatively fast. And, more importantly, the algorithm does not require small timestep discretisation to ensure convergence. 
Thus, the algorithm is particularly suited for simulating long-time trajectories. 
In general, a performance comparison between the Gillespie algorithm and other methods for stochastic open quantum system simulation depends on the specific system at hand. As will be detailed in the Discussion section, the Gillespie algorithm is particularly well suited for a very large number of trajectories on relatively small quantum systems. In contrast, it necessitates comparatively large computational resources for large, many-body simulations.

This paper is structured as follows.
After a brief overview of the SS-QJU method in Sec.~\ref{sec:QJU}, we detail our algorithm in Sec.~\ref{sec:alg}, and then compare it to alternative approaches in Sec.~\ref{sec:bench}. Finally, illustrative examples are provided in Sec.~\ref{sec:examples}. 
For the latter, we illustrate our method in comparison with the standard QJU solvers from QuTip.
An implementation of our algorithm in Mathematica is publicly available in the Melt library~\cite{implementation_Melt}, and an implementation in Julia in Ref.~\cite{implementation_Julia}.

\section{Step-by-step Quantum Jump Unravelling}
\label{sec:QJU}

In this section, we briefly review and motivate the step-by-step method for the simulation of the QJU. 
For an infinitesimal time $dt$, the evolution of Eq.~\eqref{M} can be decomposed as 
\begin{equation}\label{evo_dt}
    \rho(t+dt) = e^{\mathcal{L} dt} \rho(t) = \sum_{k=0}^r M_k \rho M_k^\dagger,
\end{equation}
where $M_k$ are Kraus operators: for $k=1,\ldots,r$ they read $M_k = \sqrt{dt} L_k$, while for $k=0$ we have $M_0 = 1- i dt H_e$, where 
\begin{align}\label{He}
    H_e =& H - \frac{i}{2}  J, \qquad \text{ with} \\\label{J}J :=& \sum_k L_k^\dagger L_k,
\end{align}
is a non-Hermitian Hamiltonian. 
The Kraus operators satisfy 
$M_0^\dagger M_0 + \sum_{k=1}^r M_k^\dagger M_k = \mathbb{I} + \mathcal{O}(dt)^2$.

The decomposition in~\eqref{evo_dt} motivates the interpretation in terms of quantum jumps.
For each timestep, one $M_k$ is chosen with probability 
$p_k = \tr(M_k \rho M_k^\dagger)$, and the system is updated to
\begin{equation}\label{update_rule}
    \rho \to \frac{M_k \rho M_k^\dagger}{\tr(M_k \rho M_k^\dagger)}.
\end{equation}
If $k=1,\ldots,r$, we say a jump occurred in ``channel'' $k$. 
Otherwise, if $k=0$, no jump occurred. 
The latter is much more likely since $p_0 \sim \mathcal{O}(1)$ while $p_k \sim \mathcal{O}(dt)$.
This yields a \emph{quantum trajectory}; i.e., a stochastic evolution of the system consisting of a few abrupt jumps, connected by a no-jump (smooth) evolution described by $M_0$~\cite{Wiseman_2009,Rouchon_2022,Qutip}.

The quantum trajectory is described by a set of outcomes $(k_1,k_2,\ldots)$, corresponding to the randomly chosen operators $M_k$ at each time step. 
One can attribute the outcomes with $k=1,\ldots,r$ to a ``click'' in a detector, while $k=0$ represents no click. 
The solution of Eq.~\eqref{M}, $\rho(t) = e^{\mathcal{L} t} \rho(0)$, is called the \emph{unconditional evolution}, while the stochastic trajectory generated by the update rule~\eqref{update_rule} is said to be \emph{conditional}, since it is conditioned on the specific set of outcomes $(k_1,k_2,\ldots)$.
The unconditional evolution is recovered by averaging the conditional evolution over multiple trajectories.

\section{The Gillespie algorithm}
\label{sec:alg}

A typical quantum trajectory, obtained for instance with the SS-QJU method outlined in the previous section, will have the form 
\begin{equation*}
    000000001000000000200000000100000001\ldots,
\end{equation*}
consisting of many $0$s, interspersed by rare jumps (in this case in channels labelled as 1 and 2). 
Clearly, it is simpler to just label the trajectories by the random times between jumps $T_i$, and the channel $k_i$ that each jump went into. That is, the quantum trajectory can instead be described by the set of outcomes
\begin{equation}\label{gillespie_trajectories}
    (T_1,k_1), (T_2,k_2), (T_3,k_3),\ldots.
\end{equation}
The Gillespie algorithm~\cite{Gillespie_1976, Gillespie_1977} avoids integrating the update rule (Eq.~\eqref{update_rule}) over infinitesimal timesteps, by instead sampling the pairs $(T_n,k_n)$ with correct probabilities.
This is done using the waiting-time distribution (WTD) which can be built as follows. 
If no jump occurs for a time $T$, at which point a jump into channel $k$ happens, then the state of the system is updated, up to a normalisation, to: 
\begin{equation}\label{gillespie_update}
    \rho \to L_k e^{-i H_e T} \rho e^{i H_e^\dagger T} L_k^\dagger.
\end{equation}
Note here that the term $e^{-i H_e T}$ results from the sequential application of the no-jump Kraus operator, in the infinitesimal limit: 
\begin{align}
\lim\limits_{n\to \infty} (\mathbb{I}-i H_e t/n)^n = e^{-i H_e t}. 
\end{align}
The normalisation factor for the rhs of Eq.~\eqref{gillespie_update} yields precisely the WTD~\cite{brandes_2008,Landi2023} 
\begin{equation}\label{WTD_basic}
    W(t,k|\rho) = \tr\big\{ L_k^\dagger L_k e^{-i H_e t} \rho e^{i H_e^\dagger t}\big\}.
\end{equation}
If we can sample over the WTD, then we may evolve the system directly from one jump to the next, rather than having to integrate the dynamics step-by-step. In some rare cases, it may be possible to obtain an analytic expression for the WTD. It is then possible to sample from it by using the efficient inversion method~\cite{breuer2007}. More frequently, however, an analytic expression for the WTD is not known, and the method of choice is the Monte Carlo Wavefunction (MCW)~\cite{Dalibard_1992, Molmer_1996, Dum_1992} when the conservation of purity is guaranteed on a trajectory, or the SS-QJU when this is not the case (e.g., for systems undergoing partial monitoring). The Gillespie method that we present here is an efficient alternative to the latter, whose speed-up is mostly due to the possibility of pre-computing some of the most numerically expensive quantities.

In order to arrive at a sampling procedure for Eq.~\eqref{WTD_basic} we first decompose it into two parts as $W(t,k|\rho) = P(k|t,\rho) W(t|\rho)$, where 
\begin{align}\label{WTD_J}
    \begin{split}
        W(t|\rho) =& \sum_k W(t,k|\rho)\\ =& \tr\big\{ J e^{-i H_e t} \rho e^{i H_e^\dagger t}\big\},
    \end{split}
\end{align}
with $J = \sum_k L_k^\dagger L_k$, as in Eq.~\eqref{J}. 
Here it is assumed that 
\begin{equation}
    \int_0^\infty dt ~W(t|\rho) = 1,\qquad  \forall\rho.
    \label{eq:normalisation_wtd}
\end{equation}
Physically, this is the case when a jump to some channel has to eventually occur for any initial state, i.e., dark subspaces are not allowed. We discuss below the situations when this might not be the case.
Letting $\tilde{\rho}=e^{-i H_e t} \rho e^{i H_e^\dagger t}$, the remaining factor is given by
\begin{equation}\label{P_k_t_rho}
    P(k|t,\rho) = \frac{\tr\big\{ L_k^\dagger L_k \tilde{\rho}\big\}}{\tr\big\{ J \tilde{\rho}\big\}}.
\end{equation}
The sampling is thus split in two: first, we sample a time $T$ using Eq.~\eqref{WTD_J},  evolving the system from $\rho\to \tilde{\rho}$. 
Next, we choose which channel the system will jump to by sampling over Eq.~\eqref{P_k_t_rho}.
This last part is comparatively straightforward, as there are only $r$ options. 
The most challenging technical issue is how to sample from Eq.~\eqref{WTD_J}.

To overcome this challenge, we write 
\begin{align}
    W(t|\rho) =& \tr\{Q(t) \rho\big\}\text{, with}\\
    \label{W_t_Q_t_basic}
    Q(t) =& e^{i H_e^\dagger t} J e^{-i H_e t}.
\end{align}
We can pre-compute $Q(t)$ for a set of times ${\tt ts}$. 
This set should be sufficiently fine-grained to resolve the fine structure in $W(t|\rho)$, and should go up to a sufficiently large time to ensure that all possible jump times are taken into account. However, as will become clear from the examples below, the timestep does not have to be infinitesimally small. Even comparably large timesteps will yield a valid probability distribution and hence a non-diverging evolution; in effect, it represents a coarse-graining of the waiting times.
For each state $\rho$, we then compute a list $W(t|\rho) = \tr\{Q(t) \rho\big\}$ for $t\in {\tt ts}$, and sample one element $T\in {\tt ts}$ from it.
A pseudocode implementation of our algorithm is provided at the end, labelled Algorithm~\ref{alg:Gillespie_core}. 

{\bf Pure states:} Initial pure states remain pure throughout the evolution along  any trajectory, provided that all the existing jump channels are monitored. This results in a significant advantage in memory requirements, and in speed. All relevant formulas above continue to hold provided we replace $\rho$ with $|\psi\rangle\langle \psi|$:
\begin{align}
    W(t|\psi) &= \langle \psi| Q(t) |\psi\rangle, 
    \\[0.4cm]
    P(k|t,\psi) &= \frac{\langle \tilde{\psi} | L_k^\dagger L_k |\tilde{\psi}\rangle}{\langle \tilde{\psi} | J |\tilde{\psi}\rangle},
    \\[0.4cm]
    |\psi\rangle &\to L_k e^{-i H_e T} |\psi\rangle,
\end{align}
where $|\tilde{\psi}\rangle = e^{-i H_e T}|\psi\rangle$. 
An implementation of the Gillespie algorithm for the simpler case of pure states is available in Julia~\cite{implementation_Julia}. We also note that any mixed state can be written as an ensemble of pure states (e.g. in the eigenstate decomposition). If all the jump channels are monitored one can then compute the evolution of mixed states using the pure state evolution of their decomposition; this is possible because trajectories remain pure under full, fine-grained monitoring. More generally, allowing for imperfect monitoring, a full-fledged mixed-state implementation of the Gillespie algorithm is unavoidable.

{\bf Renewal processes:} For this important subclass we have~\cite{brandes_2008}
\begin{equation}\label{renewal}
   \frac{L_k \rho L_k^\dagger}{\tr(L_k \rho L_k^\dagger)} =  \sigma_k,
\end{equation}
for states $\sigma_k$ which are independent of $\rho$. 
Hence, the state after the jump is always fully reset to a specific $\sigma_k$, depending only on the jump channel.
In this case, there are only a finite number of WTDs $W(t|\sigma_k)$.
They can therefore all be pre-computed, greatly speeding up the entire process. 
Simulating renewal processes is therefore extremely efficient with the proposed method.

{\bf Sampling in-between jumps:} The Gillespie algorithm yields the state of the system at random times after each jump, according to the update rule~\eqref{gillespie_update}.
For many applications, this suffices. 
In other cases, however, one also requires the state at arbitrary times in between jumps. Only a slight additional step is required to obtain such intermediate states.
Consider the update map~\eqref{gillespie_update}, for a certain time $T$ between jumps. 
The state at any intermediate time $t<T$ before the next jump is then simply
\begin{equation}
    \frac{e^{-i H_e t} \rho e^{i H_e^\dagger t}}{\tr\big\{ e^{-i H_e t} \rho e^{i H_e^\dagger t}\big\}}, \qquad t < T. 
\end{equation}
Computing this Hamiltonian evolution does not require infinitesimal timesteps. 
Instead, one can proceed as in the usual Trotter-decomposition for unitary dynamics \cite{Nielsen2002, Georgescu2014}. 
Suppose one wishes to obtain the state at steps $\Delta t$, which does not have to be small. Define $V = e^{-i H_e \Delta t}$, then simply iterate $\rho \to V\rho V^\dagger$.
A pseudocode implementation is presented as Algorithm \ref{alg:Gillespie_filling} below.

{\bf Partial monitoring:} We can generalise the above to the case where only a subset $\mathcal{M}$ of the jump operators $L_1, \ldots, L_r$ are monitored. 
Define the jump super-operators
\begin{equation}
    \mathcal{J}_k \rho = L_k \rho L_k^\dagger, 
    \qquad k \in \mathcal{M},
\end{equation}
as well as the no-jump super-operator $\mathcal{L}_0 = \mathcal{L} - \sum_{k\in\mathcal{M}} \mathcal{J}_k$ (If all channels are monitored then $\mathcal{L}_0 \rho = -i (H_e \rho - \rho H_e^\dagger)$, with $H_e$ defined in Eq.~\eqref{He}). 
Eq.~\eqref{WTD_J} is then replaced by 
\begin{align}\label{WTD_L0}
    W(t|\rho) =& \tr\big\{ J e^{\mathcal{L}_0 t} (\rho)\big\},\quad \text{with}\\ J =& \sum_{k\in \mathcal{M}} L_k^\dagger L_k.
\end{align}
Similarly, Eq.~\eqref{P_k_t_rho} becomes
\begin{align}
    P(k|t,\rho) =& \frac{\tr\big\{ L_k^\dagger L_k \tilde{\rho}\big\}}{\sum_q \tr\big\{ L_q^\dagger L_q\tilde{\rho}\big\}}, \quad\text{ with}\\[0.2cm] \tilde{\rho} =& e^{\mathcal{L}_0 t} \rho.
\end{align}
In this case, one must pre-compute $Q(t) = e^{\mathcal{L}_0^\dagger t} (J) $, with $\mathcal{L}_0^\dagger$ the adjoint super-operator of $\mathcal{L}_0$.

{\bf Classical master equation:} The original Gillespie algorithm~\cite{Gillespie_1976,Gillespie_1977} was developed for classical (Pauli) rate equations. This is actually a particular case of the present quantum version, corresponding to Eq.~\eqref{M} with $H=0$ and jump operators $L_{ij} = \sqrt{W_{ij}} |i\rangle\langle j|$, describing jumps between (orthonormal) basis states $|i\rangle$, and $|j\neq i\rangle$, with transition rate $W_{ij}$.
For this case, Eq.~\eqref{J} reduces to 
$H_e = - \tfrac{i}{2} J$, with 
\begin{equation}
\begin{split}
    J =& \sum_{i\neq j} L_{ij}^\dagger L_{ij} = \sum_{i\neq j} W_{ij} |j\rangle\langle j|\\ =& \sum_j R_j |j\rangle\langle j|,
    \end{split}
\end{equation}
where $R_j = \sum_{i\neq j} W_{ij}$ is the \emph{escape rate} for the system to leave state $|j\rangle$. 
As $[H_e, J]=0$, now Eq.~\eqref{WTD_J} simplifies to 
\begin{equation}
\begin{split}
    W(t|\rho) =& \tr\big\{ J e^{-J t} \rho\big\} \\=& \sum_j R_j e^{-R_j t} \langle j|\rho|j\rangle.
    \end{split}
\end{equation}
The process is renewal (Eq.~\eqref{renewal} is satisfied).  
The state at the previous jump   will  be $\rho = |i\rangle\langle i|$, for some state $|i\rangle$. 
Hence, the WTD becomes 
\begin{equation}
    W(t|i) = R_i e^{- R_i t},
\end{equation}
which is an exponential distribution with rate $R_i$. 
Sampling it is thus trivial\footnote{If $r$ is uniform between 0 and 1, then $-\frac{1}{R_i} \ln r$ is exponentially distributed with rate $R_i$.}, which is the reason why the Gillespie algorithm is so efficient in the classical context.

{\bf Dark subspaces:} Throughout, we assumed that $\mathcal{L}_0$ is invertible, which ensures that a jump must always happen. 
In some systems this might not be the case. 
The most direct way to check this is to see if $\mathcal{L}_0$ has any zero eigenvalues (i.e., a non-empty nullspace). 
Alternatively, one can also compute the no-jump probability 
\begin{equation}
    P_{\rm no}(t|\rho) = \tr\big\{ e^{\mathcal{L}_0 t} \rho\big\}.
\end{equation}
If $P_{\rm no}(\infty|\rho) = \lim_{t\to\infty} P_{\rm no}(t|\rho)=0$, then a jump must always occur. 
This can  sometimes be the case even if $\mathcal{L}_0$ is not invertible, depending on the initial state $\rho$~\cite{Landi2023c}. 
One may verify that since $\mathcal{L}_0 = \mathcal{L}-J$ and since $\mathcal{L}$ is traceless, it follows that $P_{\rm no}$ is related to $W(t|\rho)$ in Eq.~\eqref{WTD_L0} according to
\begin{equation}
    W(t|\rho) = - \frac{d  P_{\rm no}(t|\rho)}{dt}.
\end{equation}
Thus, the normalization condition becomes $\int_0^\infty dt W(t|\rho) = 1-P_{\rm no}(\infty|\rho)$. 

{\bf Choice of the time parameters:} We observe that, by construction, our algorithm will always retrieve properly normalised states. If the time step $dt$ is chosen too large, however, the simulation may not be able to capture the short-time details of the Hamiltonian evolution. Nonetheless, $dt$ does not have to be infinitesimal, as in the SS-QJU method, since the convergence of a differential equation is not at stake here. Instead, it suffices that it is small enough to ensure the WTD is smooth. The final sampling time $t_f$ must be chosen large enough to guarantee that waiting times larger than $t_f$ happen with negligible probability. 

\section{Example applications}\label{sec:examples}
To better illustrate the advantages and shortcomings of the Gillespie algorithm, in this section we present a range of applications for various systems of physical interest. 
In all simulations below, we used the Julia implementation of the Gillespie algorithm, available at \cite{implementation_Julia}. 
The results will be compared with SS-QJU simulations performed using QuTip. The first examples showcase our approach for simple physical systems. The last one applies the Gillespie algorithm to the case of partial monitoring, where its advantage is the most significant.

\subsection{Single-qubit resonant fluorescence}
We considered first a very well-studied quantum optics problem, one-qubit resonant fluorescence \cite{Wiseman_2009, Carmichael_1989}. A single qubit evolves with Hamiltonian ($\hbar = 1$)
\begin{equation}
    H = \Delta \sigma^z + \Omega \sigma^x,
\end{equation}
where $\Delta$ represents a detuning term and $\Omega$ is the Rabi frequency due to coupling with an external electromagnetic field. Given a leak rate $\gamma$ for the qubit, its time evolution is described by the Lindbladian super-operator
\begin{equation}
\begin{split}
    \mathcal{L}\rho =& -i[H, \rho]  + \frac{\gamma}{2} \left(2 \sigma^- \rho \sigma^+ - \{\sigma^+\sigma^-, \rho\}\right),
\end{split}
\end{equation}
where $\sigma^\pm = \sigma^x \pm i \sigma^y$. The process is renewal, and the form of the waiting time distribution is analytically known \cite{Carmichael_1989} and reads, in the case $\Delta = 0$
\begin{equation}
    W(t) = \frac{16 \gamma  \Omega ^2 e^{-\frac{\gamma  t}{2}} \sinh ^2\left(\frac{1}{4} t
   \sqrt{\gamma ^2-16 \Omega ^2}\right)}{\gamma ^2-16 \Omega ^2}, 
\end{equation}
where the initial state is $\rho = |\downarrow\rangle\langle \downarrow|$ is the state after the previous jump. 

\begin{figure}[htbp]
    \includegraphics[width=\columnwidth]{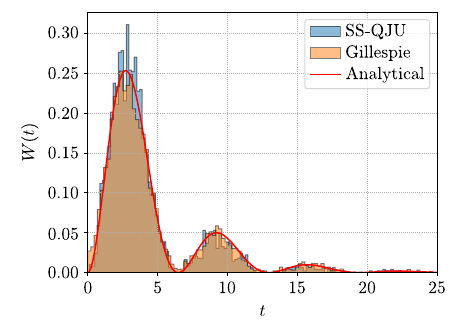}  
    \caption{Single-qubit resonant fluorescence: simulated waiting time distribution with 250 trajectories, with SS-QJU and with the Gillespie algorithm. Parameters: $\Delta = 0$, $\Omega = \gamma = 0.5$; initial state $\ket{0}$ For the SS-QJU, $dt = 0.01$, for the Gillespie algorithm {\tt ts} has intervals of 0.01 and is limited to a final time of 100. 
    }\label{fig:resonant_fluorescence}
\end{figure}

The results of the simulation, both with the Gillespie algorithm and SS-QJU, are shown in Fig.~\ref{fig:resonant_fluorescence}. We also plot the analytical results on top. The agreement with the analytics and with the SS-QJU simulations provides a good sanity check for the correctness of the method.

\subsection{Double qubit non-renewal process}
\begin{figure}[h!]
    \includegraphics[width=0.99\columnwidth]{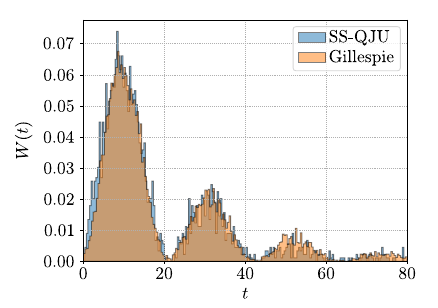}
    \caption{Double-qubit system: simulated waiting time distribution with SS-QJU and the Gillespie algorithm. Parameters: $\Omega = 3$, $\gamma = 0.1$, $g=0.3$; initial state $\ket{0}\otimes\ket{0}$. In SS-QJU, $dt=0.01$, for the Gillespie algorithm {\tt ts} has intervals of 0.01 and is limited to a final time of 1000.}
    \label{fig:double_qubit}
\end{figure}

In order to discuss the behaviour of the Gillespie algorithm when relaxing the renewal hypothesis, we considered a slightly more complex physical system. Two qubits are coupled with an exchange interaction (coupling intensity $g$); one of them is driven by an external electromagnetic field (Rabi frequency $\Omega$), while the other can decay with rate $\gamma$. The Hamiltonian and the Lindbladian are:
\begin{equation}
    H = \Omega \sigma_1^x + g \left(\sigma_1^+ \sigma_2^- + \sigma_1^- \sigma_2^+\right);
\end{equation}
\begin{equation}
\begin{split}
    \mathcal{L}\rho = &  -i[H,\rho] \\
    & + \frac{\gamma}{2} \left(2 \sigma^-_2\rho\sigma^+_2 - \{\sigma^+_2 \sigma^-_2, \rho\}\right).
\end{split}
\end{equation}
The waiting time distributions are shown in Fig.~\ref{fig:double_qubit}, again for both methods.
In both this and the previous example, we have found the Gillespie algorithm to be highly efficient, owing to the low dimensionality of the Hilbert space (which implies a low memory overhead).

\subsection{Mesoscopic charge qubit under continuous quantum measurement}
As a further application, we considered the model of mesoscopic charge qubit presented in Ref.~\cite{Goan_2001_1, Wiseman_2009, Goan_2001}. The system is given by two coupled quantum dots, with Hamiltonian
\begin{equation}
    H = \sum_{j=1}^2 \omega_j \sigma_j^+\sigma_j^- + \Omega(\sigma_1^+ \sigma_2^- + \sigma_1^-\sigma_2^+),
\end{equation}
where $\omega_1$ and $\omega_2$ represent the electron annihilation/creation energy for the respective dot, and $\Omega$ sets the coupling strength. The Lindbladian is given by
\begin{equation}\label{Example_3_Liouvillian}
    \mathcal{L}\rho = - i [H,\rho] + \mathcal{D}[\mathcal{T} + \chi n_1]\rho,
\end{equation}
where $\mathcal{T}$ and $\chi$ represent the tunnelling amplitudes between the two dots.

\begin{figure}[htbp]
    \includegraphics[width=\columnwidth]{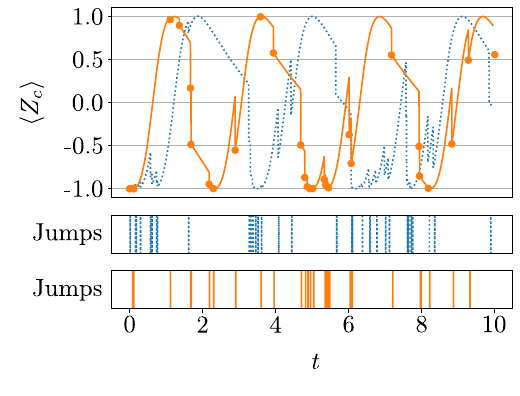}
    \caption{Single-trajectory simulated expectation value of the population difference for the mesoscopic charge model, computed both with SS-QJU (dotted blue line) and the Gillespie algorithm (solid orange line). For the latter, the expectation value of the observable $Z_c$ on the state right after the jump is represented with a solid circle. In the lower panels, the corresponding jump times are shown, as computed by the two methods. Parameters: $\omega_1 = \omega_2 = \Omega = \mathcal{T}=\chi = 1$; initial state $\ket{1}\otimes\ket{0}$. For SS-QJU, $dt=0.001$, for the Gillespie algorithm {\tt ts} has intervals of 0.001 and is limited to a final time of 10.}
    \label{fig:Goan_Milburn}
\end{figure}

We reproduced the stochastic trajectory for the expectation value of the population-difference operator $Z_c$ (Fig.~2 in Ref.~\cite{Goan_2001}) using the Gillespie algorithm in Fig.~\ref{fig:Goan_Milburn}, together with the stochastic sequence of jump times. This simulation shows the possibility to interpolate the states between the jump times computed using the Gillespie algorithm. 

Note that the Gillespie algorithm does not give a relevant advantage when only a small number of trajectories is considered. In some cases, due to the pre-computation stage, it can actually take longer than time-discretisation-based algorithms. Instead, the power of the Gillespie algorithm we propose comes to bear when a large number of trajectories is taken into account.
For this specific model, this can be quite advantageous because quite often one has $\mathcal{T} \gg \chi$ in Eq.~\eqref{Example_3_Liouvillian}. This means that many jumps will occur, but most of these will be uninformative. 
It therefore becomes relevant to acquire the statistics over a very large number of jumps, which is where the Gillespie method becomes particularly powerful.

\subsection{Kerr model}
\label{sect:Kerr_model}
\begin{figure}[h!]
    \includegraphics[width=\columnwidth]{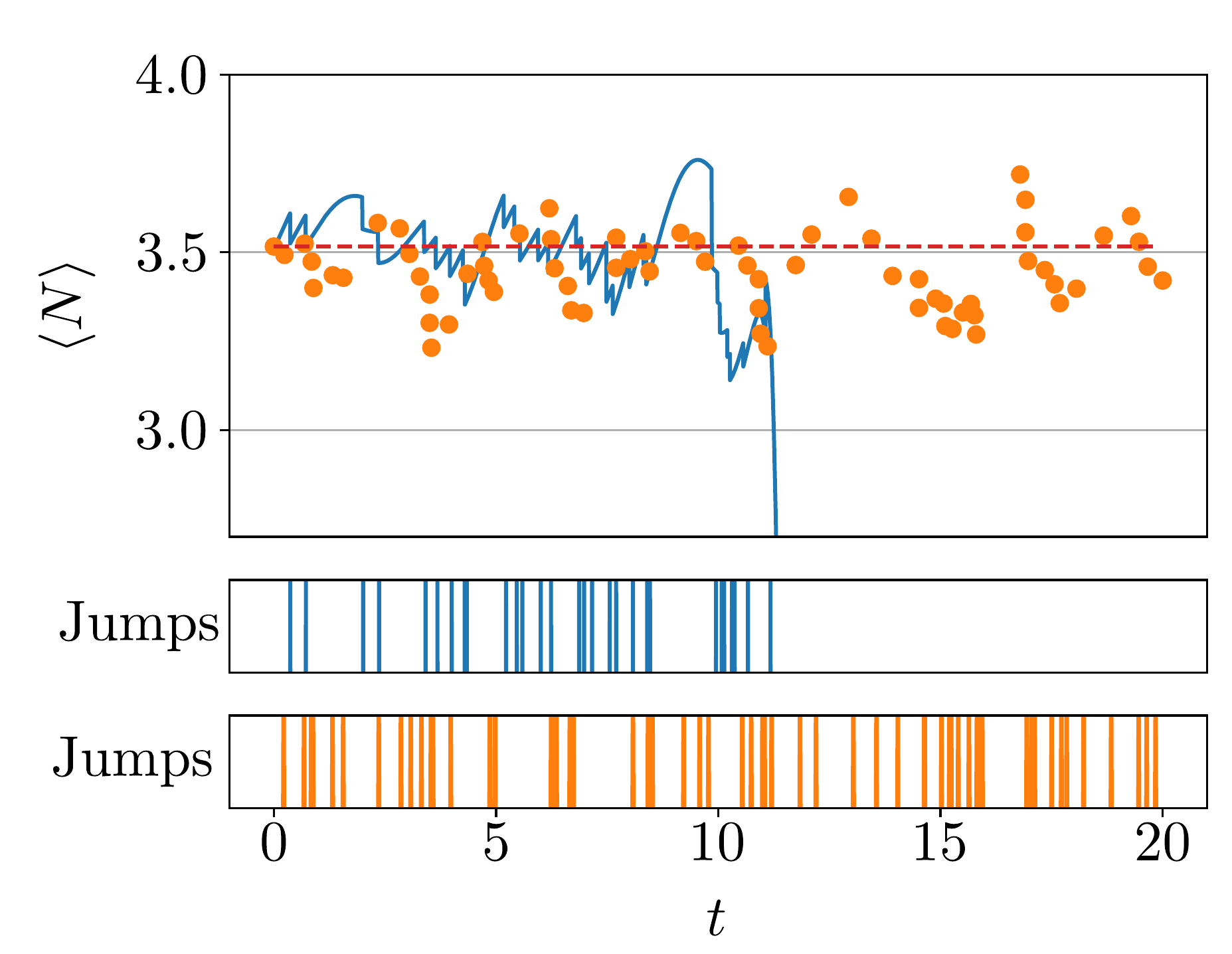}
    \caption{Single-trajectory simulated expectation value of the number operator for the Kerr model presented in the main text, computed with both SS-QJU (dotted blue line) and the Gillespie algorithm (solid orange circles). The expectation value of $N$ on the steady state is represented by the red dashed line. The lower panels show the jump times obtained for the trajectories at the top, corresponding to each method, respectively. Note how, for long times, SS-QJU does not converge due to the error accumulation effect detailed in the main text, while the Gillespie algorithm does. Parameters: $\Delta=1.5$, $U=0.05$, $F=3.27$, $\gamma=1$. For SS-QJU we used $dt = 0.0001$, while for the Gillespie algorithm {\tt ts} has intervals of 0.01 and is limited to a final time of 50. The Hilbert space size is cut at 30 Fock states, and the initial state is the steady state of the system.}
    \label{fig:Kerr_model}
\end{figure}

Finally, we consider the application of the Gillespie method to the simulation of the  Kerr model~\cite{Drummond1980} described by the Hamiltonian
\begin{equation}
    H = \Delta a^\dagger a + \frac{U}{2} a^\dagger a^\dagger a a + F^* a^\dagger + F a,
\end{equation}
where $a$ represents a bosonic annihilation operator, $\Delta$ is a detuning parameter, $U$ represents the intensity of the non-linear coupling, and $F$ is the driving strength of an external laser. The system undergoes a jump-like evolution, described by the Lindbladian super-operator
\begin{equation}
    \mathcal{L}\rho = -i \left[H, \rho\right] + \frac{\gamma}{2}\left(2 a \rho a^\dagger - a^\dagger a \rho - \rho a^\dagger a\right),
\end{equation}
where $\gamma$ is the decay rate. 
Quantum trajectories for this model have recently been obtained experimentally in~\cite{Fink_2018}, and our approach can be used to directly compare with that experiment. 
In particular, we simulated the behaviour of the expectation value of the number operator $a^\dagger a$ as a function of time, both with SS-QJU and our Gillespie algorithm. The results for a single trajectory are shown in Fig.~\ref{fig:Kerr_model}. Contrary to the previous example, in this case we did not fill the gaps between the jumps in the case of the Gillespie algorithm: the expectation value of the number operator is then known only for the state right after the jump.  This is compensated by a significant gain in performances on the side of the Gillespie algorithm: due to pre-computation of the heaviest parts of the simulation, obtaining data for a large number of trajectories can be much faster than with other methods.

\subsection{Three Level Maser}
\begin{figure*}[htbp]
    \centering
    \includegraphics[width=\textwidth]{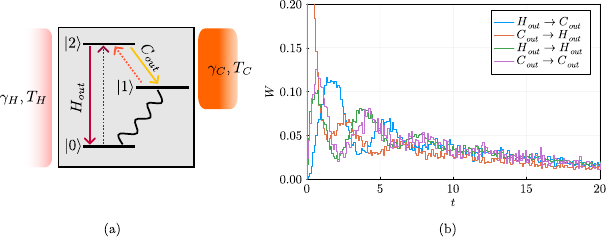}
    \caption{(a) Schematic of the three level maser model. The jumps represented by solid lines are monitored, the channels represented by dotted lines are unmonitored. The wavy line denotes the Hamiltonian evolution. (b) Relevant waiting time distributions for all possible combinations of the two monitored jumps in the three level model. Parameters: $\gamma_H = 0.1$, $\gamma_C = 2$, $\Delta = 0$, $\Omega = 0.8$, $n_H$ = 5, $n_C = 0.025$, $dt = 0.01$, final evolution time 100, $2\times \, 10^4$ trajectories.}
    \label{fig:three_level_maser}
\end{figure*}

To illustrate the application of the Gillespie algorithm in the case of partial monitoring, we consider the three level maser model~\cite{Scovil_1959, Kalaee_2021, VanVu_2022}. The maser, represented in Fig.~\ref{fig:three_level_maser}a, consists of three energy levels, $\ket{0}$, $\ket{1}$ and $\ket{2}$, evolving according to the Hamiltonian
\begin{equation}
\begin{split}
    H & = \omega_0 \sigma_{00} + \omega_1 \sigma_{11} + \omega_2 \sigma_{22} \\ &
    + \varepsilon \left(e^{i(\Delta + \omega_1 - \omega_0)t}\sigma_{01} + \text{h.c.}\right),
\end{split}
\end{equation}
where $\omega_j$ represents the energy of the level $\ket{j}$, and $\sigma_{ij} = \ketbra{j}{i}$, $\Delta$ is the detuning parameter, and $\varepsilon$ the driving intensity. The system is coupled to two thermal baths at different temperatures $T_H$ and $T_C$, with respective coupling strengths $\gamma_H$ and $\gamma_C$. In an appropriate rotating frame~\cite{Kalaee_2021}, the Hamiltonian can be rewritten as
\begin{equation}
    H^{\text{(rf)}} = - \Delta \sigma_{11} + \varepsilon\left(\sigma_{01} + \sigma_{10}\right),
\end{equation}
and the interactions with the baths take the form of four jump operators
\begin{equation}
\begin{split}
    & L_{H,\text{out}} = \sqrt{\gamma_H (n_H + 1)} \sigma_{20}; \\
    & L_{H,\text{in}} = \sqrt{\gamma_H n_H} \sigma_{02}; \\
    & L_{C,\text{out}} = \sqrt{\gamma_C (n_C + 1)} \sigma_{21}; \\
    & L_{C,\text{in}} = \sqrt{\gamma_C n_C} \sigma_{12},
\end{split}
\end{equation}
where $n_H$ and $n_C$ are the bosonic average occupation numbers of the hot and the cold bath, respectively.

To showcase the use of the Gillespie algorithm, we consider the three level maser system undergoing partial monitoring. As represented in Fig.~\ref{fig:three_level_maser}a, we assume that the experimenter is only able to access signals corresponding to emissions from the system, but not the opposite. That is, the monitored jump channels are $H_{\text{out}}$ and $C_{\text{out}}$. 

In particular, we compute all the four possible waiting time distributions, represented in Fig.~\ref{fig:three_level_maser}b. Since purity is not conserved in systems undergoing partial monitoring, if one was to obtain the same plot using the MCW method, this would require exponentially (with time) more many trajectories. It is in the 
cases of partial monitoring and channel merging, indeed, that the power of density-matrix-based methods is unleashed. 

\section{Discussion}
\label{sec:bench}
As outlined above, two categories of simulation techniques for quantum jump trajectories are commonly in use; on one hand, SS-QJU splits the evolution into small timesteps, and for each of them determines whether an emission will probabilistically take place. MCW, in contrast, exploits the norm reduction of the conditional no-jump evolved state to sample the survival probability. MCW requires purity-preserving evolutions; this is the case for the finest unravelings of a GKSL equation, but does not hold for coarser unravelings, corresponding to e.g. partial monitoring and channel merging.

In this section, we critically discuss the Gillespie algorithm, highlighting its strenghts and shortcomings. The natural point of comparison is SS-QJU, as MCW is not directly applicable for coarser unravelings.

As common in physics simulations, we note a trade-off between memory footprint and computational time. The Gillespie algorithm saves calculation time by pre-computing most quantities of interest (in particular the vector $Q(t)$), which are trajectory-independent, before the actual simulation. This works very well when the dimension of the considered Hilbert space is relatively small (e.g., on the order of $N\sim100$ or less). On the contrary, for high-dimensional systems, the memory required to store all the pre-computed quantities may pose a challenge. Similarly, our algorithm requires, in Eq.~\eqref{gillespie_update}, the exponentiation of the effective Hamiltonian; this passage can be computationally costly for large systems.

Once the relevant quantities have been pre-computed, the simulation of each individual trajectory with the Gillespie algorithm is very fast. In fact, computation is required only in correspondence with jumps; no further pre-calculation is required for the evolution between jumps. Considering the interplay between pre-computation and proper simulation time, the Gillespie algorithm has a low marginal computational cost per trajectory and thus works well for simulating large numbers of trajectories. Thereby, the initial disadvantage due to pre-computation is offset by the advantage in each individual trajectory simulation. An interesting step in achieving even better performance would be the parallelisation of the trajectory evolution, such that each thread separately samples a subset of the required trajectories. Analogously the pre-computation step can also easily be parallelised. Parallelisation would further improve the execution time, but not reduce the memory footprint.

In many cases, the Hamiltonian dynamics happens on a different (usually faster) timescale than the jumps. A simulation method that follows the evolution step-wise therefore has to employ a very small time step to follow the fast Hamiltonian oscillations; however, during most simulation steps no jumps will occur, resulting in wasted computation time compared to the `fast-forward' simulation logic underlying the Gillespie algorithm. The algorithm therefore performs well when in the regime of rare jumps.

The Gillespie algorithm does not natively compute the system state at all times, but only to those immediately after a jump. In this, it mimics the information available to an experimenter, who has no access to the full state history, but only to the measurement record in terms of jumps and channels. While it is possible to fill the gaps between two jumps (see Algorithm~\ref{alg:Gillespie_filling}) to recover the state from a Gillespie algorithm-based simulation, this partially undoes some of the algorithm's advantage. Hence, the algorithm is particularly suited when the measurement record rather than full knowledge of the state at arbitrary times is required.

The Gillespie algorithm is based on the initial pre-computation of the function $Q(t)$. Under time-dependent driving this precomputed function would not represent the actual dynamics. Without further generalisation, e.g. to periodic driving, the Gillespie algorithm is thus not suitable for the simulation of systems undergoing explicitly time-dependent driving.

Algorithms that have to follow the Hamiltonian evolution step-wise tend to be very sensitive to the simulation time-step $dt$, which has to be small enough to follow the oscillations of the coherent dynamics. If $dt$ is chosen too large, the simulation will lose convergence after some time. The Gillespie algorithm does not need to follow the entire Hamiltonian evolution, therefore avoiding this convergence issue. It therefore exhibits good numerical stability. An example of this behaviour is shown in Fig.~\ref{fig:Kerr_model}.

\section*{Conclusions}
In this work, we have presented a newly developed algorithm to efficiently simulate quantum jump trajectories: the Gillespie algorithm, named after the classical algorithm that inspired it. It follows a different logic from other algorithms for the same purpose in that it allows leaping directly from one quantum jump to the next, without having to go through many steps during which no jumps take place. In contrast to MCW simulation, which follows a similar logic -- coherent evolution interspersed by jumps according to the delay or waiting time distribution~\cite{Cohen_1986,Dalibard_1992, Dum_1992, Molmer_1996} --, the Gillespie method can be used when the purity of individual trajectories in not preserved, e.g. due to partial monitoring or channel merging. The algorithm is therefore particularly suitable for those cases in which there is a significant difference in timescales between very fast Hamiltonian dynamics and much slower jump dynamics, and where only partial information is gained during the jumps. We presented examples of applications of the Gillespie algorithm to different systems of interest, highlighting its advantageous points with respect to alternative simulation techniques, such as those based on time discretisation or MCW methods. Adding to these two approaches to the stochastic simulation of open quantum systems, the quantum Gillespie algorithm thus adds an additional simulation paradigm to the open quantum systems toolkit.

\section*{Acknowledgements}
The research conducted in this publication was funded by the Irish Research Council under grant numbers IRCLA/2022/3922 and GOIPG/2022/2321. The authors thank Michael Kewming and Simon Milz for helpful comments, respectively, on QuTiP and on partial monitoring.

\bibliography{bibliography}

\begin{algorithm*}
\caption{Gillespie evolution for mixed states and/or partial monitoring}\label{alg:Gillespie_core}
\KwIn{Hamiltonian $H$;}
list of monitored jumps Kraus operators [$L$]\;
list of unmonitored jumps Kraus operators [$S$]\;
initial state density matrix $\rho_0$\;
final time $t_f$\;
timestep $dt$\;
number of trajectories\;
\Precompute\
$J=\sum_k L_k^\dagger L_k$, $\tilde{J}=\vectorize(J)$\;
$\tilde{\mathcal{L}_0} = -i\mathbb{I}\otimes H+iH^T\otimes\mathbb{I} +\sum_l \left(S_l^* \otimes S_l - \frac{1}{2} \mathbb{I}\otimes S_l^\dagger S_l - \frac{1}{2}(S^\dagger_lS_l)^T\otimes\mathbb{I}\right) + \sum_k \left(-\frac{1}{2} \mathbb{I}\otimes L_k^\dagger L_k - \frac{1}{2} (L_k^\dagger L_k)^T \otimes \mathbb{I}\right)$\;
$\tilde{\mathcal{L}}_0^\dagger = i \mathbb{I}\otimes H - i H^T\otimes \mathbb{I} + \sum_l \left(S_l^T\otimes S_l^\dagger - \frac{1}{2}(S_l^\dagger S_l)^T \otimes \mathbb{I} - \frac{1}{2} \mathbb{I}\otimes (S^\dagger_l S_l)\right) + \sum_k \left(-\frac{1}{2}(L_k^\dagger L_k)^T\otimes \mathbb{I} - \frac{1}{2} \mathbb{I}\otimes L_k^\dagger L_k\right)$\;
Let [$V$] be the list of the no-jump evolution superoperators\;
Let $Q_s$[$t$] be the list of the non-state-dependent parts of the WTD\;

\For{all times $t < t_f$}{
    $V[t] = \exp(\tilde{\mathcal{L}}_0 t)$\;
    $Q_s[t] = \unvectorize(\exp(\tilde{\mathcal{L}}_0^\dagger t)\tilde{J})$\;
}
\For{all trajectories}{
    Fix initial state $\rho = \rho_0$, $\tau=0$\;
    \While{$\tau < t_f$}{
        Let [$W$] be the list to contain the WTD\;
        \For{every value $Q$ in [$Q_s$]}{
            $W = \tr[Q\rho]$
        }
        Sample a time $T$ from $W$\;
        $\tau += T$\;
        $\rho = \unvectorize(V[T]\vectorize{\rho})$
        Let [$\rho$] be the list of weights of the different (monitored) jump channels\;
        \For{$L$ in [$L$]}{
            $p[L] = \tr[L^\dagger L\rho]$\;
        }
        Sample a jump channel $k$ from the weights distribution [$p$]\;
        $\rho = \frac{L_k\rho L_k^\dagger}{\tr[L_k\rho L_k^\dagger]}$
    }
}
\KwOut{[$(t_j, k_j, \rho_j)$] vectors of results and states after jumps for all trajectories, [$V$], [t range] list of times at which $V$ is known}
\end{algorithm*}

\begin{algorithm*}
\caption{Finding the state at fixed times along a trajectory}\label{alg:Gillespie_filling}
\KwIn{[$t^*$] list of times at which $V$ is known}
[$t$] list of times at which the state has to be computed\;
[$V$] list of evolution superoperators at times specified in [$t^*$]\;
[($j_j, k_j, \rho_j)$] vector of results and states for a single trajectory\;
Set to simplify the notation a final jump at time $+\infty$\;
\For{each jump $(t_j, k_j, \rho_j)$ along the trajectory}{
    Determine the set [$\bar{t}$] of times such that $t_j \leq \bar{t}< t_{j+1}$\;
    \For{each $\bar{t}$ in [$\bar{t}$]}{
        Find the closest time $t_{cl}$ in [$t^*$] to $\bar{t}-t_j$\;
        $\rho = \frac{\unvectorize(V[t_{cl}]\vectorize(\rho))}{\tr[\unvectorize(V[t_{cl}]\vectorize(\rho))]}$
    }
}
\KwOut{the vector of $\rho$ states at all times in [$t$]}
\end{algorithm*}
\end{document}